# The process of 3D-printed skull models for the anatomy education

**Zhen Shen[1*] • Yong Yao[2*] • Yi Xie[1,3] • Chao Guo[1] • Xiuqin Shang[1] • Xisong Dong[4] • Yuqing Li[1,5] • Zhouxian Pan[6] • Shi Chen[7,8] • Hui Pan[7,8,9] • Gang Xiong[10]**

**Abstract**

*Objective* The 3D printed medical models can come from virtual digital resources, like CT scanning. Nevertheless, the accuracy of CT scanning technology is limited, which is 1mm. In this situation, the collected data is not exactly the same as the real structure and there might be some errors causing the print to fail. This study presents a common and practical way to process the skull data to make the structures correctly. And then we make a skull model through 3D printing technology, which is useful for medical students to understand the complex structure of skull.

*Materials and Methods* The skull data is collected by the CT scan. To get a corrected medical model, the computer-assisted image processing goes with the combination of five 3D manipulation tools: Mimics, 3ds Max, Geomagic, Mudbox and Meshmixer, to reconstruct the digital model and repair it. Subsequently, we utilize a low-cost desktop 3D printer, Ultimaker2, with polylactide filament (PLA) material to print the model and paint it based on the atlas.

*Result* After the restoration and repairing, we eliminate the errors and repair the model by adding the missing parts of the uploaded data within 6 hours. Then we print it and compare the model with the cadaveric skull from frontal, left, right and anterior views respectively. The printed model can show the same structures and also the details of the skull clearly and is a good alternative of the cadaveric skull.

*Conclusion* We present an available and cost-effective way to obtain a printed skull model from the original CT data, which has a considerable economic and social benefit for the medical education. The steps of the data processing can be performed easily. The cost for the 3D printed model is also low. The manipulation procedure presented in this study can be applied widely in processing skull data.

✉ Gang Xiong xionggang@casc.ac.cn

✉ Hui Pan panhui20111111@163.com

1. The State Key Laboratory for Management and Control of Complex Systems, Institute of Automation, Chinese Academy of Sciences (CASIA), Beijing 100190, China
2. Department of Neurosurgery, Peking Union Medical College Hospital (PUMCH), Chinese Academy of Medical Sciences & Peking Union Medical College (CAMS & PUMC), Beijing 100730, China
3. School of Electrical & Electronic Engineering, the University of Manchester, Manchester M13 9PL, the United Kingdom
4. Qingdao Academy of Intelligent Industries, Qingdao 266109, China
5. College of Information Science and Technology, Beijing University of Chemical Technology, Beijing 100029, China
6. Eight-year Program of Clinical Medicine, PUMCH, CAMS & PUMC, Beijing 100730, China
7. Department of Endocrinology, Endocrine Key Laboratory of Ministry of Health, PUMCH, CAMS & PUMC, Beijing 100730, China
8. National Virtual Simulation Laboratory Education Center of Medical Sciences, PUMCH, CAMS & PUMC, Beijing 100730, China
9. Department of Education, PUMCH, CAMS & PUMC, Beijing 100730, China
10. Cloud Computing Center, Chinese Academy of Sciences, Dongguan 523808, China

* These authors contributed to the work equllly and should be regarded as co-first authors.

## Introduction

The 3D printing technology is developing fast these years and it has many advantages, such as customization, rapid manufacturing, accuracy and relatively low cost. We could print as many models as we want [12].

There have been various applications on the combination of 3D printing technology and medical science [13]. Waran et al. presented a procedure of making accurate 3D spatial models of the nasal cavity, paranasal sinuses (sphenoid sinus in particular), and intrasellar/pituitary pathology, according to the data of an individual patient [4]. Lee and his teammates obtained the pathology images of the maxillofacial defect through the CT scanner and input the data into a computer to reconstruct the 3D model and evaluated the degree of defect. The process played an important part in the maxillofacial defect repair operation, which shows the advantages of the combination of CT, the 3D technology and clinical medicine application [5]. Matthew and his colleagues made use of similar methods to print the scapular osteochondroma model of a 6-year-old girl. The visualization of the model helped the doctor to analyze the pathology and the situation [6]. Debarre's team took the examples of the osteotomy for epiphyseal malunion, shoulder arthroplasty and femoral trochleoplasty to show that the 3D printing technology can help make apparent hidden or ambiguous details and make sense in orthopaedic and trauma surgery. Their work showed that the 3D printing can be applied in the reconstruction of lesion and defect area and it is better than using only the CT 3D reconstruction [7]. The Sectra Inc. offers solution for planning on 3D images improves planning of such complex cases as multiple fragment traumas. With the solution, it is easy to visualize trauma structure and diagnose a fracture. And Sectra's 3D Core solution enables manual or semi-automatic bone segmentation [14]. Nowadays, the technology of making medical models through 3D printing is developing fast. Various materials with different colors and hardness are ready to choose. The 3D model is easier to understand than the 2D image shown on the computer. In this case, the 3D printed model has become a significant tool in the surgery education and training.

The anatomy teaching plays an important role in the modern medical education. The traditional teaching methods mainly include using virtual 3D images, 2D atlases and cadaveric skulls. The virtual 3D images cannot be touched so it may be not helpful in the surgery education. The 2D atlases need the students to have a good spatial imagination. Cadaveric skulls are good. However, due to the high cost of maintaining, limited number of donors and the ethical concerns, the opportunity to touch the cadaveric skull is not enough for the education purpose [1,2,3]. New methods are desirable.

The anatomy structure of skulls has been always challenging for medical students in their surgery curriculums and study. The reason is that the complex 3D structures are strange to the students. A 3D model can go over the difficulty well and Lay a solid foundation for novices to learn surgery. Normally, the procedure of making a printed model goes from CT scanning data to 3D data and then to the printed model.

In this paper, we present a procedure of using computers and equipments to make a skull model. This procedure includes the computer-assisted image processing, which fixes the model data to make the digital model accurate and ready to print. Then we choose the proper materials for the printing. To make medical students comprehend the model better, we paint the model based on the atlas. The procedure is shown in Fig.1.

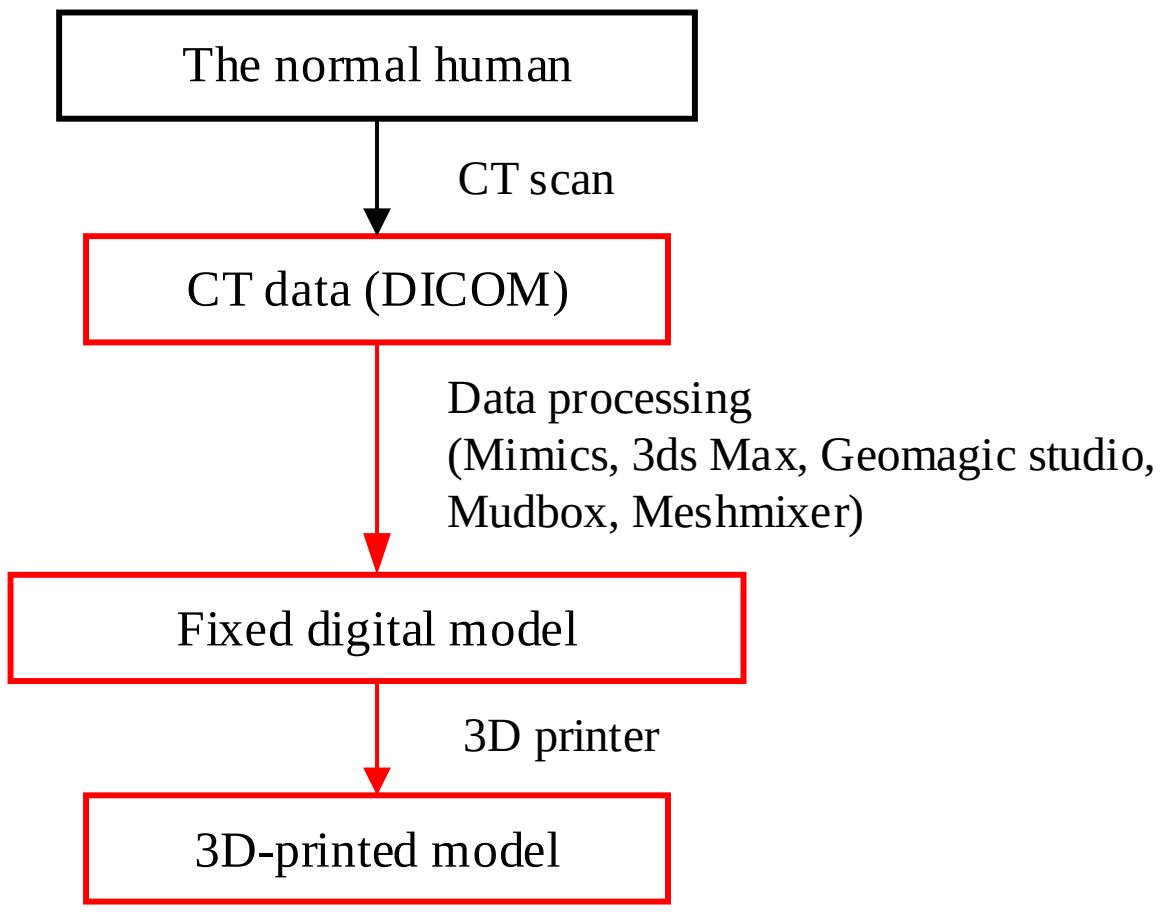

Fig.1 The model manufacture process. We collect CT data (DICOM files) and then process it with several software including Mimics, 3ds Max, Geomagic Studio, Mudbox and Meshmixer to obtain the fixed digital model. Finally, we use a 3D printer to print the model.

## Materials and methods

### Overview

We use Siemens SOMATOM Force CT scanner (horizon scan, 1mm) to scan a normal male's head to collect the data in the axial plane, with a slice width of 1 mm, a helical pitch of 0.9, and an image production interval of 1 mm.

The accurate structure of the skull model is the most significant element for a medical model. However, the procedure of collecting the digital data and format transformation may cause some mistakes or missing of the structure. In this case, repairing the digital model is the most important and difficult step during the whole process. Table 1 has summarized the main problems and corresponding solutions in this study.

Table.1 Summary of main problems and corresponding solution.

| Problems | Solution | 3D manipulation tools | Tools |
|---|---|---|---|
| File format transformation | Upload the DICOM file and save it as STL file | Mimics (17.0) | *File > New Project Wizard; File > Export > ASCII STL Files* |
| Non-skull parts | Set the thresholding value | | *Menu bar > Segmentation > Thresholding* |
| | Select the parts in 3D and delete them | | *Menu bar > Segmentation > Thresholding; Menu bar > Segment > Edit Mask in 3D > Lasso > Remove* |
| Missing structure | Paint the parts slide by slide | | *Menu bar > Segment > Edit Masks > Circle > Draw* |
| Open skull | Cut off the top of the skull | 3ds MAX (2014) | *Slice > Remove the top* |
| Wrong positions of holes | Fill the holes | Geomagic Studio (13.0) | *Fill Holes > Fill All* |
| Inappropriate distribution of grids | Analysis and optimize | | *Polygons > Repair > Remesh;*<br>*Polygons > Smooth > Relax;*<br>*Polygons > Repair > Delimate* |
| Deficient feature | Carve the specific section | Mudbox (2012) | *The toolkit* |
| Digital model checking | Review the digital model | Meshmixer | *Analysis* |

As the real procedure, the detailed operation and results are described by the order of 3D manipulation tools in the following parts.

**Mimics Procedure**

The scan data is exported to a digital imaging and communications in medicine (DICOM) file, and converted to STereoLithography (STL) file by the Mimics 17.0.

The CT result contains the entire human organs and body structure. We have to obtain the needed part of the skull by setting the thresholding value. Since not all the body structures are required during the skull anatomy education, such as the muscle, cartilaginous skeletons and some CT equipment hardware, we edit the mask and delete them.

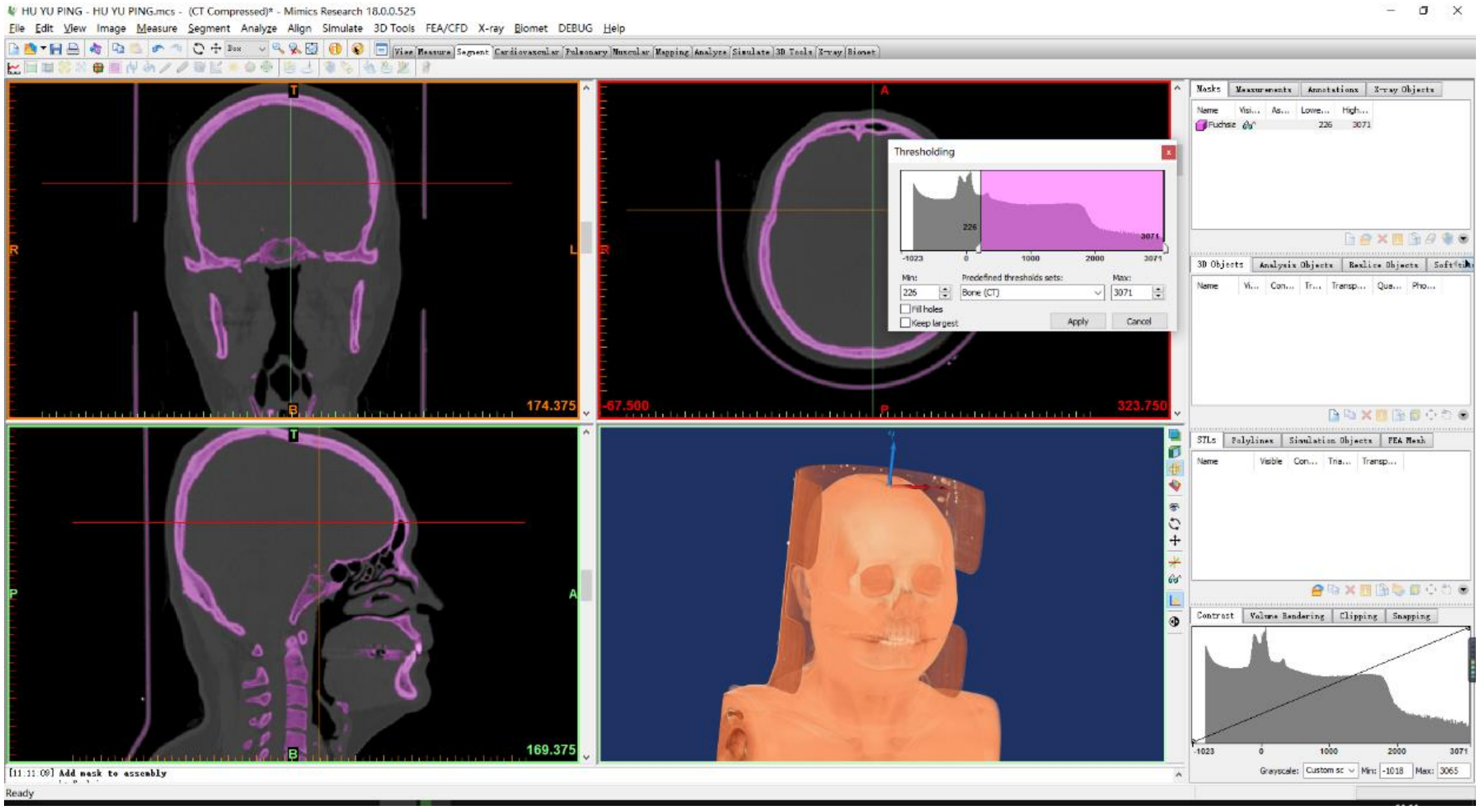

Fig.3 Set the thresholding value. When the CT scan file is uploaded, we select *Menu bar* > *Segmentation* > *Thresholding*, and, for the data, we set the minimum value as 226 (Bone(CT)) to get the required part of the skull model. The thresholding result is saved as a new mask automatically.

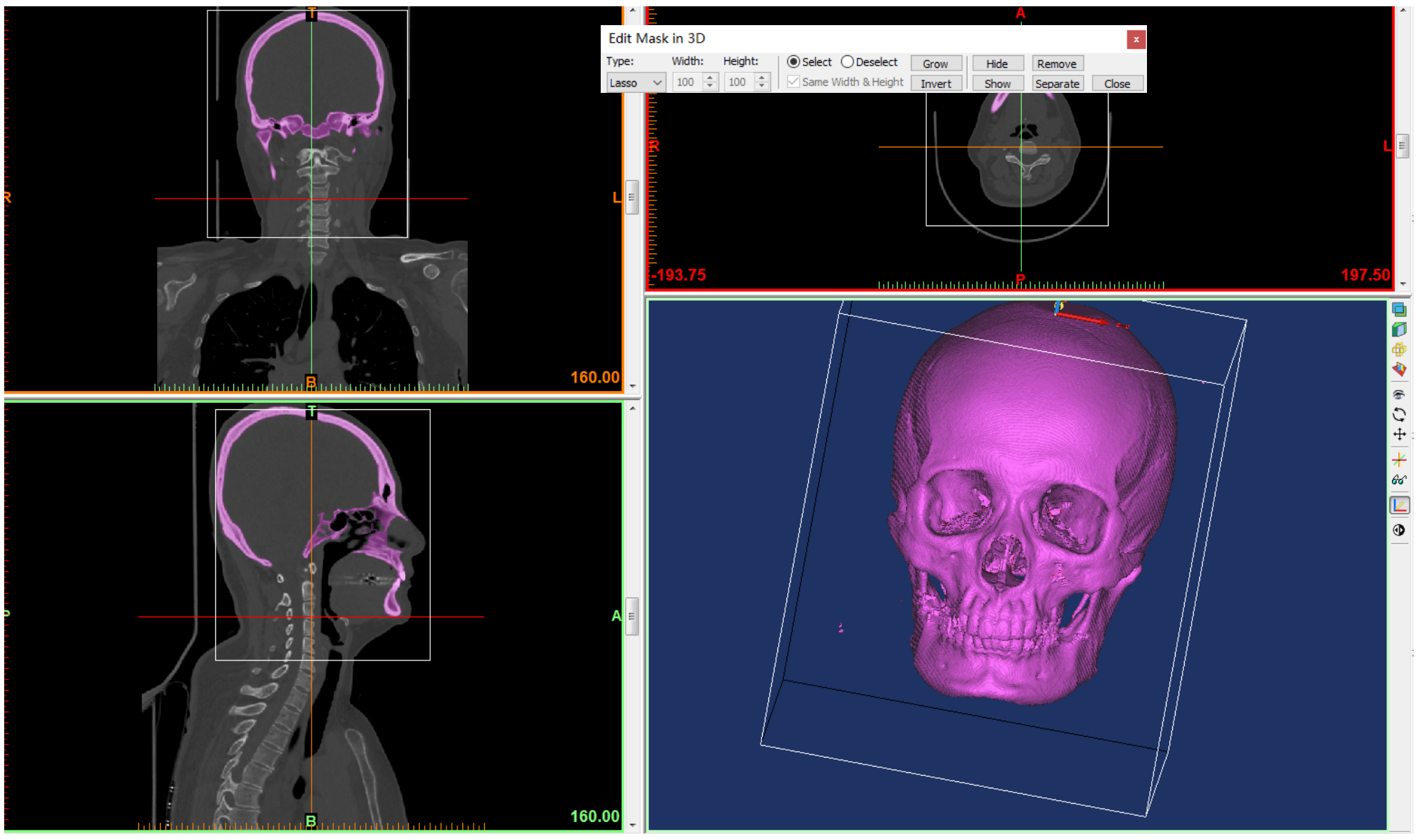

Fig.4 Delete the unwanted part of the model. We select *Menu bar* > *Segment* > *Edit Mask in 3D*, and use the "*Lasso*" to choose the unrequired parts, such as the vertebra. And then, we select "*Remove*" to delete it.

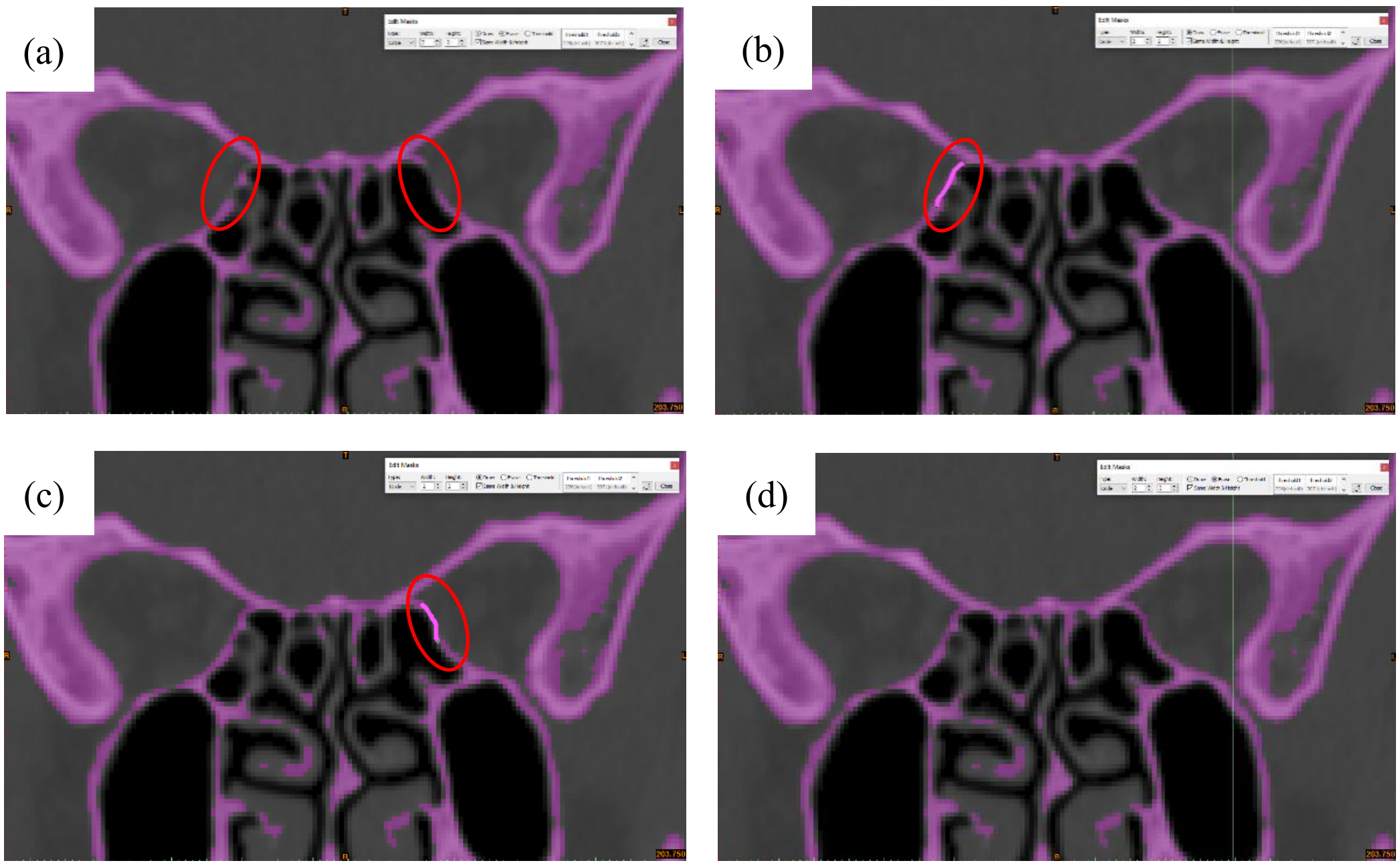

Fig.5 Repair the holes. **a** The purple parts in the red circle should be connected. The CT scanning and thresholding value setting may cause some missing parts of the digital model. **b** Using *Menu bar* > *Segment* > *Edit Masks* and selecting "*Circle*" with reasonable size and selecting "*Draw*" to connect the part on the left. **c** Using the same steps as **b** to repair the part on the right. **d** the complete slice. Since the skull is stereoscopic, we have to "*Draw*" the missing parts slice by slice.

Meanwhile, the vertebra is a solid, which is unrequired and difficult to remove in *Edit Mask* tool, we use the *Edit Mask in 3D* > *Erase* to remove them.

Therefore, the operations of adding or removing required and unrequired parts accordingly are easy and quick in the Mimics.

## 3ds Max Procedure

Since there may be several missing, damaged or excess structures after accomplishing the above procedure, we have to delete the excess parts and repair the damaged and missing parts to make sure that the model is unbroken so that we can print it. Here, we use the 3ds MAX (2014) to process the model.

In the 3ds Max, we use the "*slice*" function to delete the skull cap. (Fig.6)

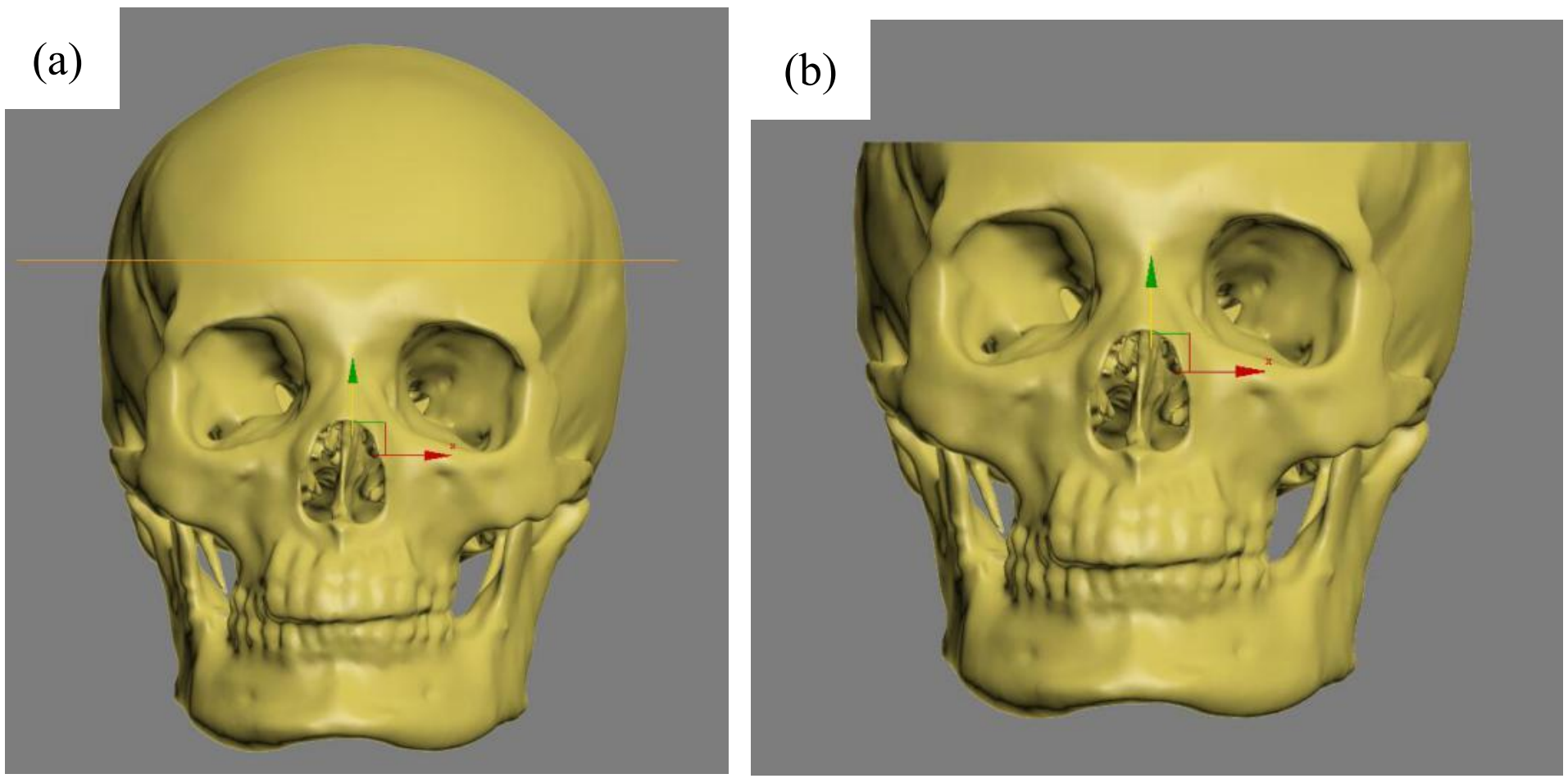

Fig.6 Cut the top part. **a** Select the "*Slice*" tool. Move the cursor to a suitable place. **b** Selecting "*Remove the top*" to delete the skull cap.

## Geomagic Studio Procedure

After removing the top skull, the skull is not an enclosure, which can cause difficulties when printing. We use Geomagic Studio (13.0), to repair it (Fig.7).

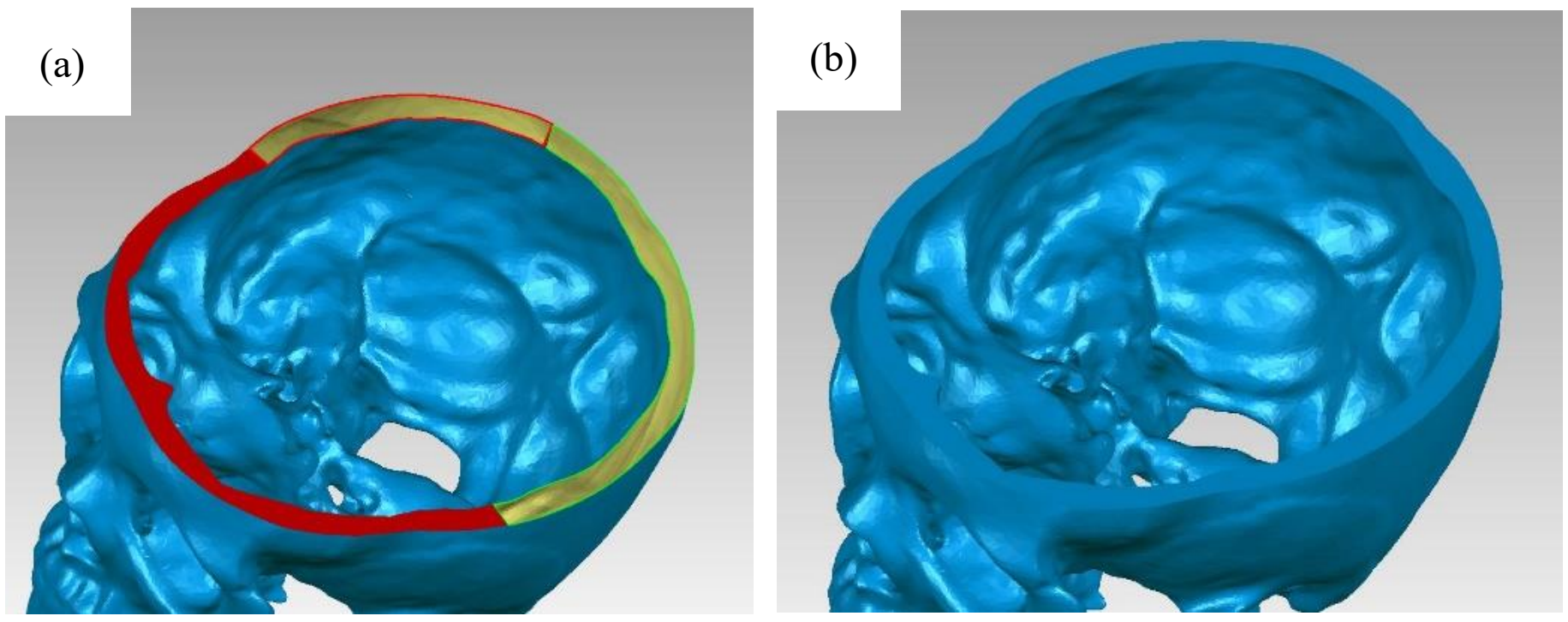

Fig.7 Repair the top. **a** Use *Fill Holes* > *Bridge* in the edge of the skull cap. Then use the *Fill Holes* > *Fill All* to repair. **b** the surface is level now.

The *Fill Holes* tool of Mimics performs well in fixing the missing structure.

It is difficult to use a single thresholding value to distinguish compact bone and cancellous bone, so there are some small holes in the surface of the skull. Then, we fix these holes (Fig.8).

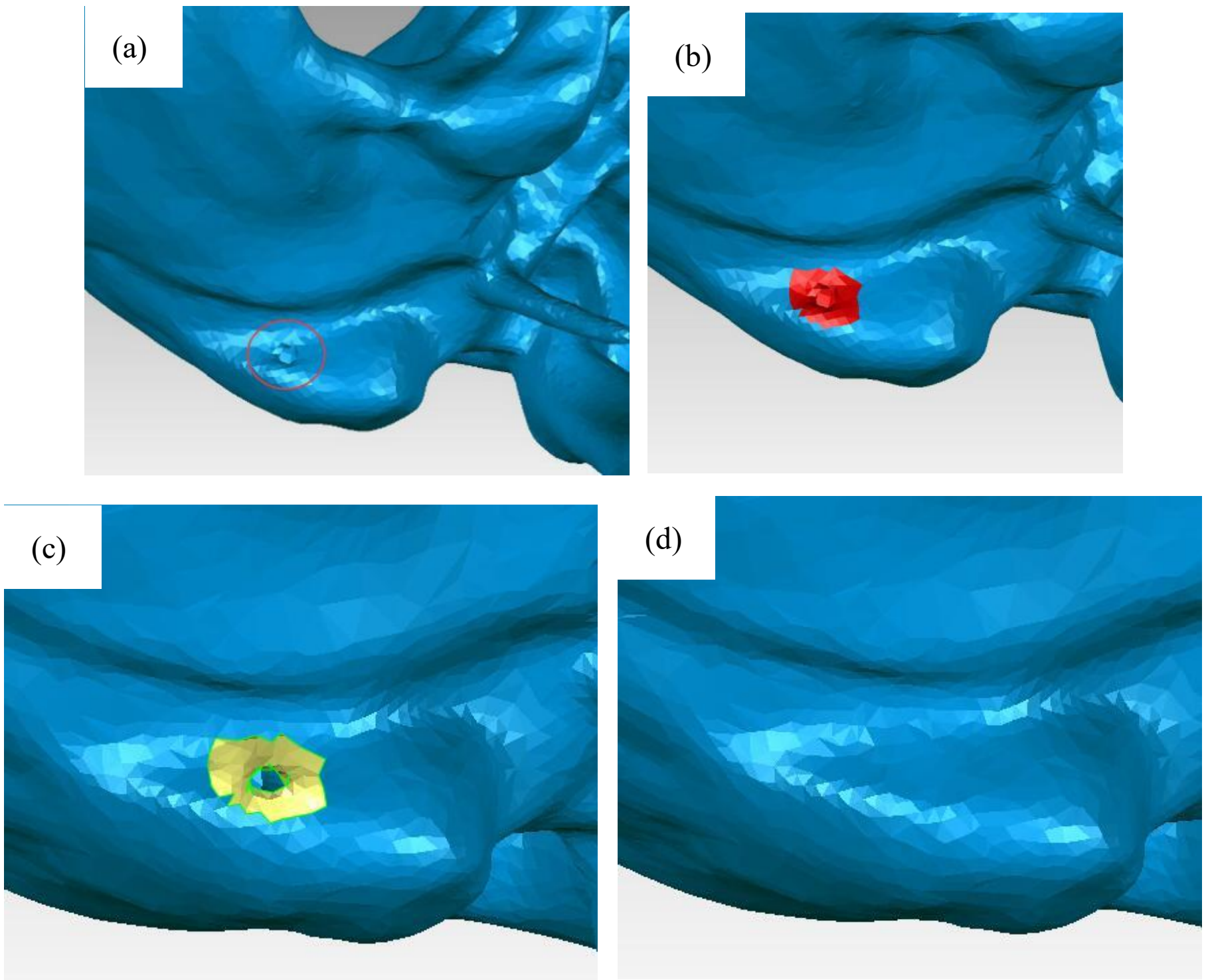

Fig.8 Repair the hole in the surface. **a** There is a hole in the red circle. **b** We select the hole and the surroundings. **c** We delete the hole and make it a bigger hole. **d** We select *Fill Holes* > *Fill All* to repair the big hole.

After zooming in the digital model, we can see the triangular facets of the current model is not reasonable, such as the largely different distribution of the meshes and the large number of the triangles in the mesh. Some triangular facets are too close, which will affect the surface quality of the printed model. Hence, we are going to use tools in *Polygons* to improve the model (Fig.9).

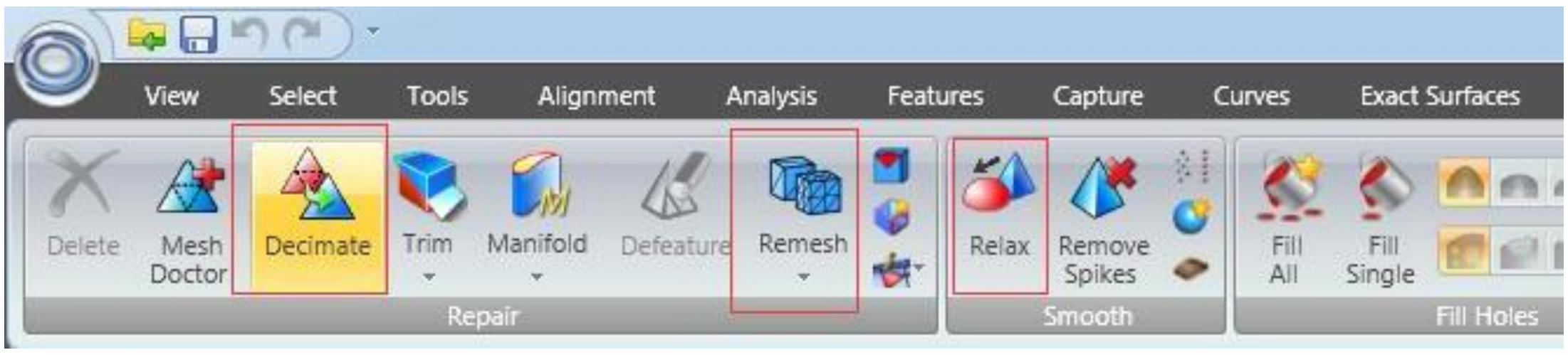

Fig.9 Improve the quality of the model. We select *Polygons* > *Repair* > *Remesh* to retriangulate a polygon mesh to produce a more uniform tessellation. Then we select *Polygons* > *Smooth* > *Relax* to smooth a polygon mesh by minimizing angles between individual polygons, selected *Polygons* > *Repair* > *Decimate* to reduce the number of triangles without reducing surface detail or color.

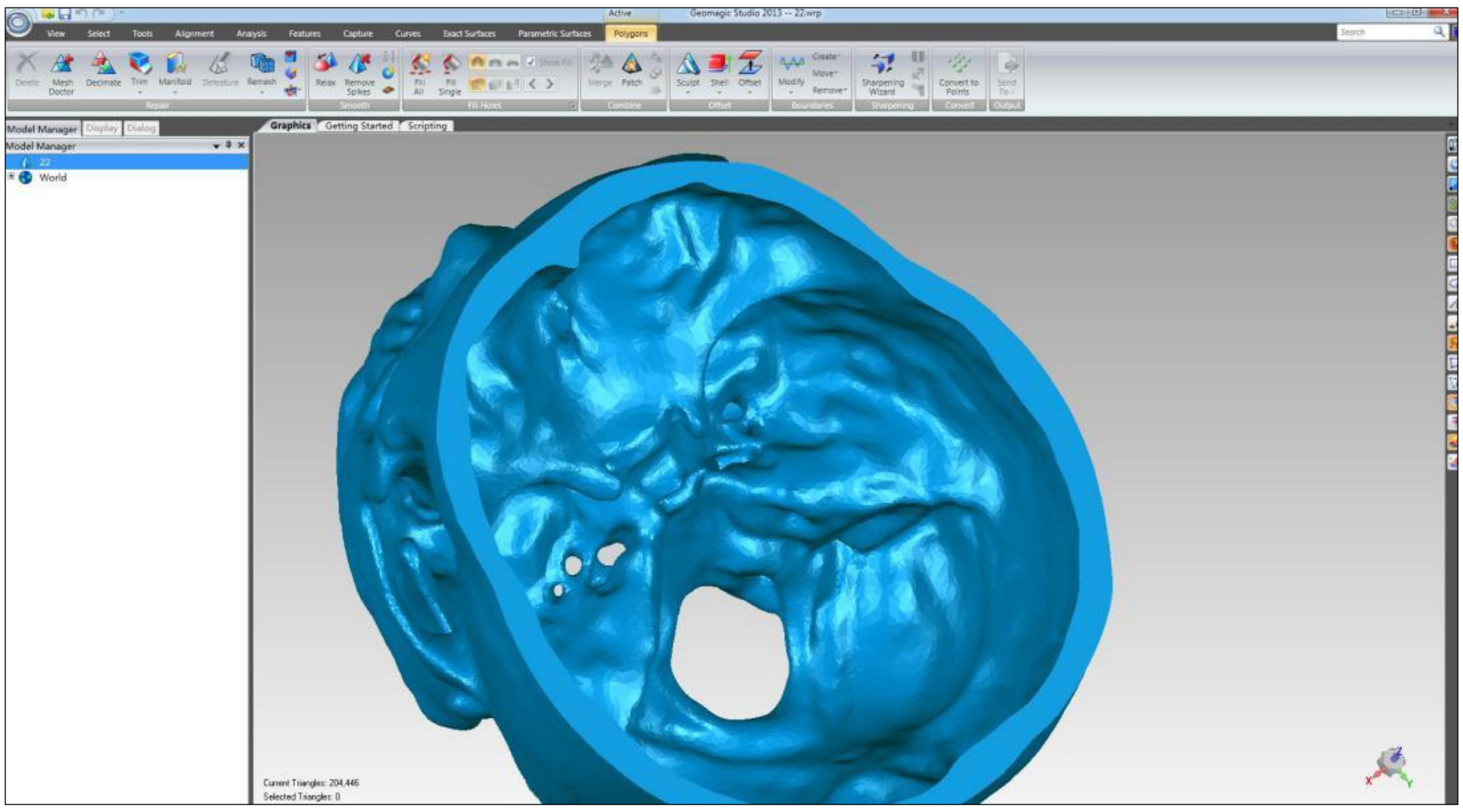

Fig.10 The model with a better quality and smooth surface.

## Mudbox Procedure

Then we export the file in the OBJ (OBJect) format and upload it into Autodesk Mudbox (2012), an easily-handled digital carving 3D manipulation tool, to further to carve some significant portion features and make the feature more obvious. (in Fig.11).

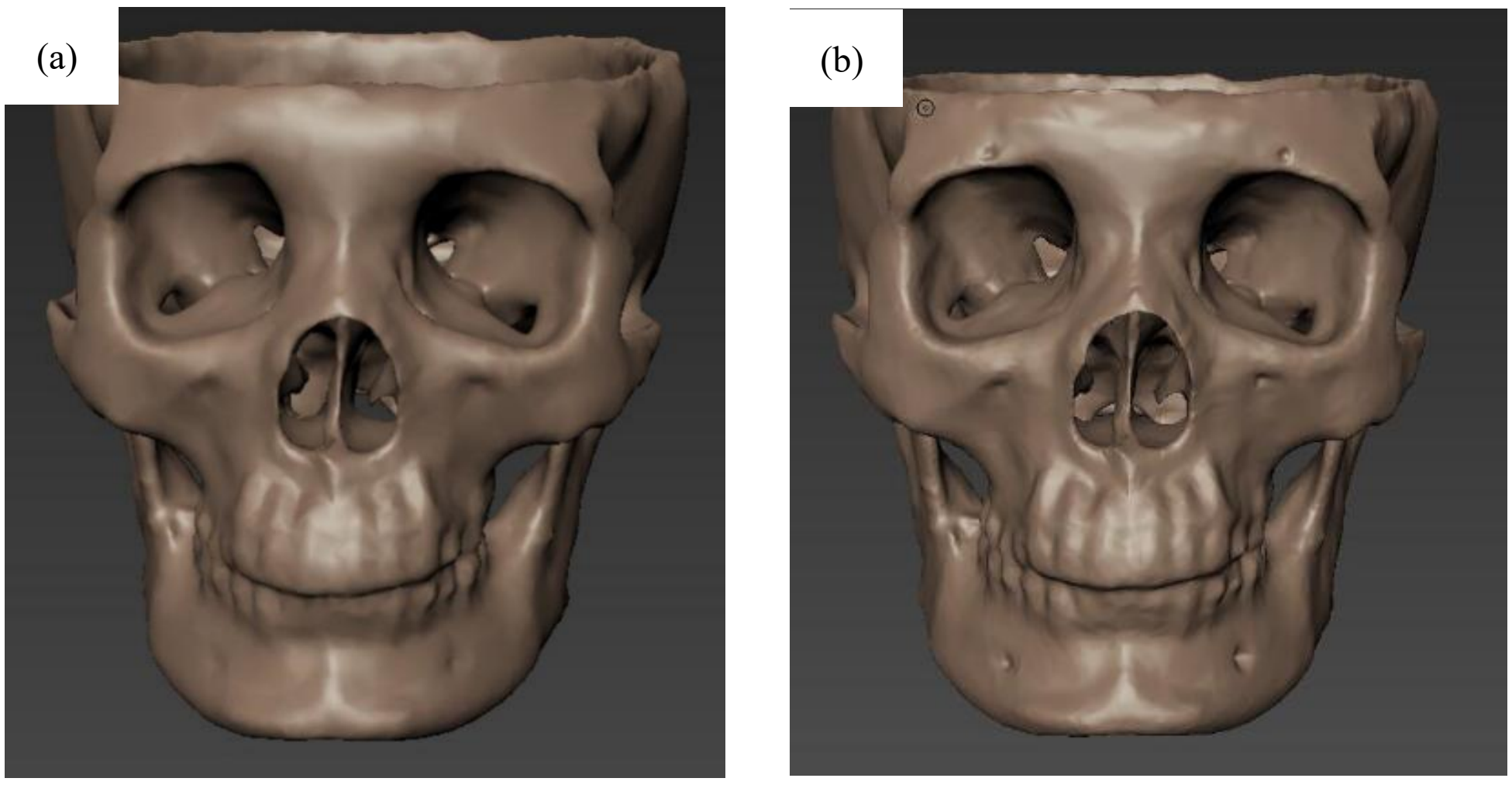

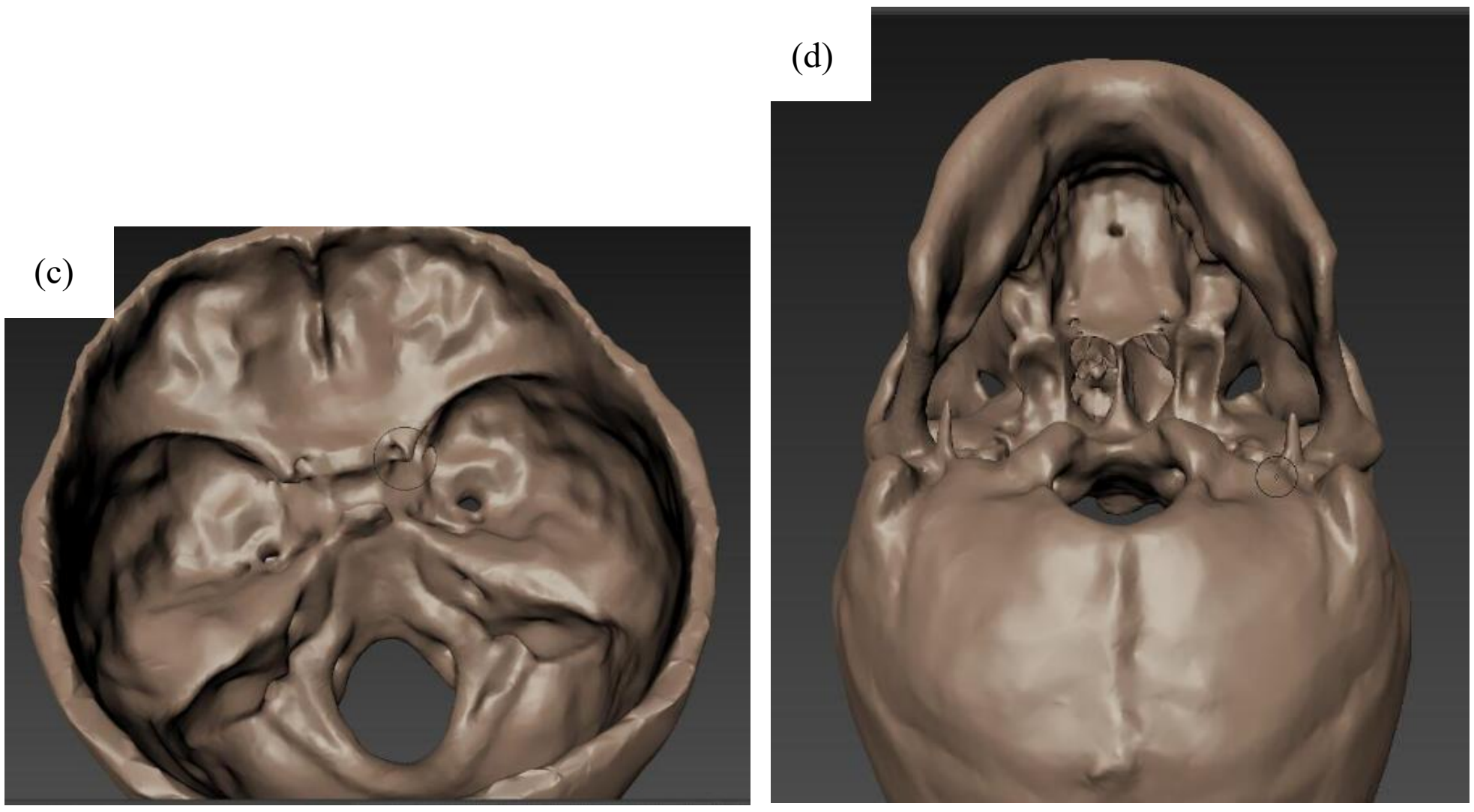

Fig.11 Make the features more obvious. **a** The original digital model before manipulation in the Mudbox. **b** We deepen the holes, such as supra-orbital foramen and infra-orbital foramen and the cheek part, which are not obvious at first. **c** we repair the anterior clinoid and make them more obvious. **d** We create two mastoid processes (Their sizes in the digital model are too small to show and to print accurately.)

### Meshmixer Procedure

Since the original data is not good enough, there may be some damaged parts in the model. In this step, we check the digital model through Meshmixer carefully before the printing procedure.

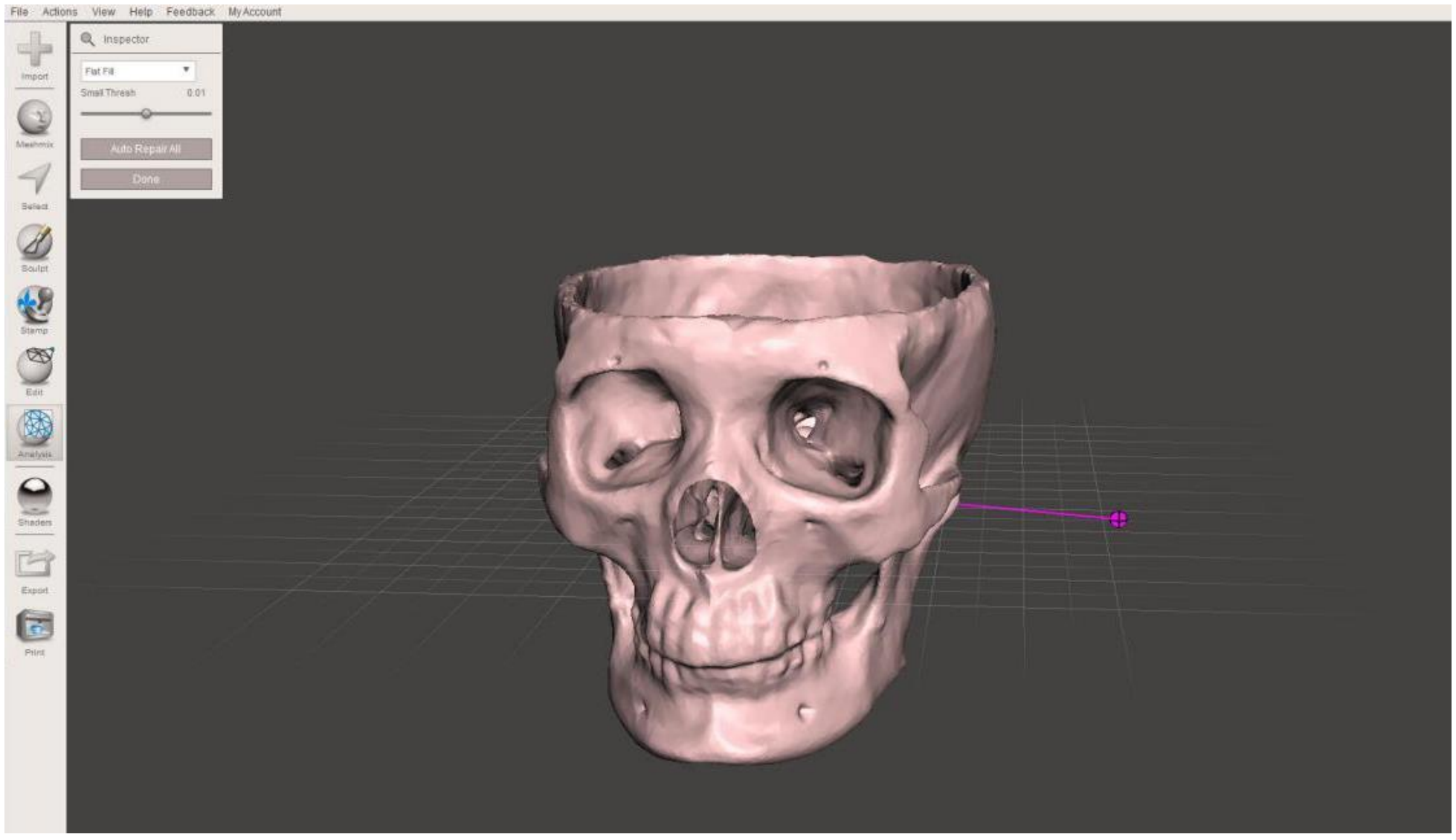

Fig.12 Review the digital model in Meshmixer and make sure that we can print it. If there are some defects remaining, we have to go back to the previous steps to repair them.

### 3D Printing Procedure

We chose the FDM (Fused Deposition Modeling) technology 3D printer, Ultimaker2 (China, 2016, $500) using white polylactic acid (PLA) materials with a diameter of 1.75 mm, and a layer thickness of 0.1 mm. We paint the model with twelve different colors propylene to distinguish different parts

based on the atlas, which can help the students learn better the structures. The printing process took 28 hours and painting took 2.5 hours. Raw materials for each skull cost approximately $14.

## Result

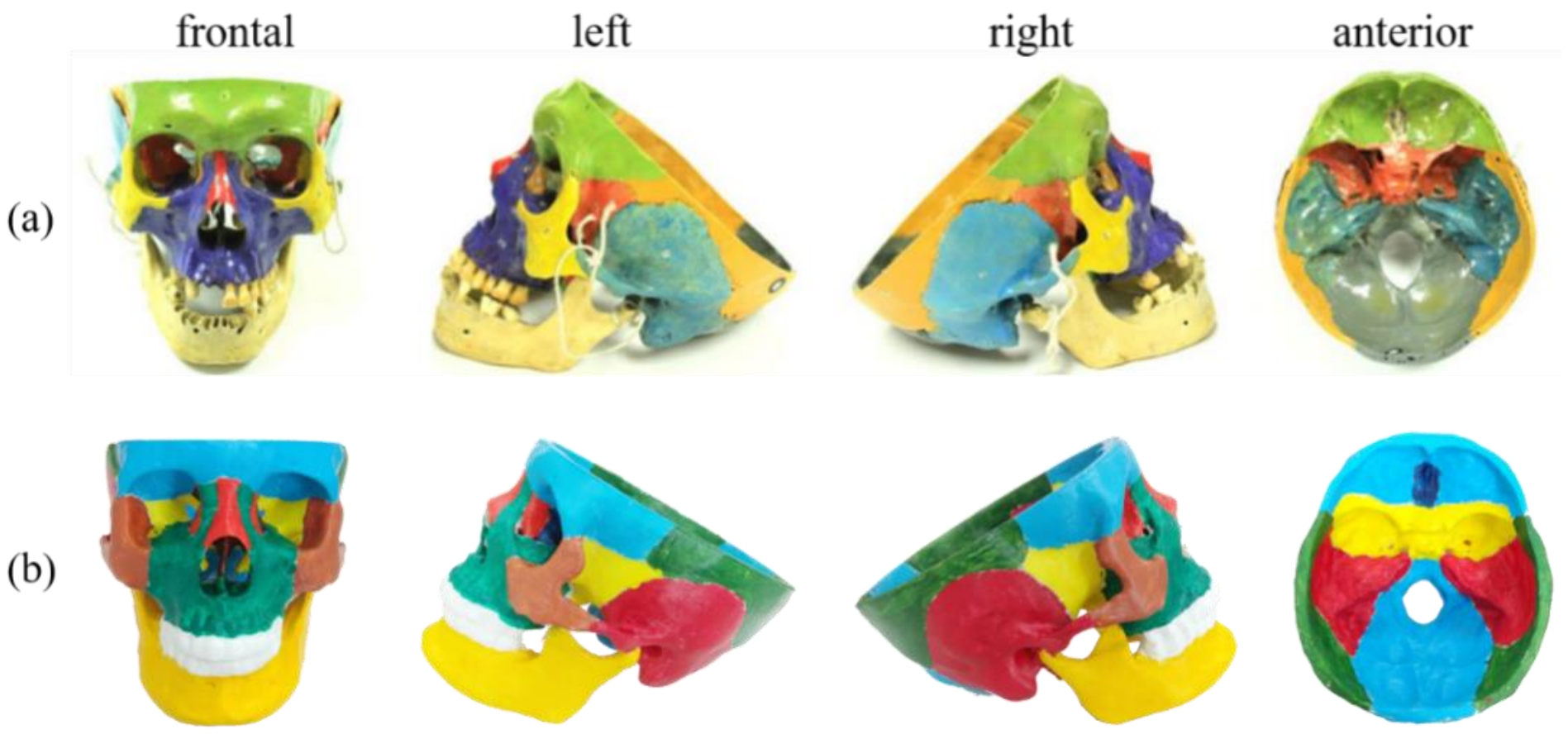

Fig.13 Photos of cadaveric skull and 3D-printed skull. **a** Cadaveric skull is showed in frontal, left, right and anterior views, respectively. **b** 3D-printed skull is showed in frontal, left, right and anterior views, respectively.

We reproduce the model 1:1, printed with twelve colors based on the atlas. We can see the completed 3D-printed skull model in Fig.13. The 3D-printed skull has the same structure as the cadaveric skull. And it is much easier to remake the 3D-printed skull.

## Discussion

The surgery needs good spatial imagination and understanding of patients' condition [9]. The current and common way to learn anatomy structure mainly relies on the cadaveric skull, specifically the dissection. However, the cadaveric skulls are not easily preserved and the opportunities of the surgeries are really rare for medical students. Besides, as the related ethics issues attracting increasing attention, there are not enough cadaveric skulls that can be used by medical students. In this situation, if we are able to obtain models showing the sophisticated anatomy structures, the surgery can be practiced and simulated repeatedly, which means a lot for the medical education. Further, it can help to increase the success probability during the real dissection surgery [11].

The 3D printing technology is useful in the medical area. The limitation to get a medical model is the cost and accuracy. The 3D printing technology shows good usages in the medical science, especially with the assistance of the medical image processing. We can obtain the model in a relative cheap, quick and convenient way by the 3D printing technology [10]. With the medical image processing, we can get the wanted model, not only the standard models but also the customized ones. Medical students, surgeons, and educational experts can do their researches or studies on these 3D-printed models.

The entire operation presented in the paper can be applied in the other 3D medical model processes. Although the CT scanning procedure and the format transformation may miss some information of the structure, or the digital structure is partly damaged or missing or too thin to print, we can follow steps shown in this paper to fix the model, by combining computer-assisted image processing with the knowledge of the anatomy structures. In the part of the data processing, we showed elaborate steps of operation. The whole procedure is summarized in Fig.14.

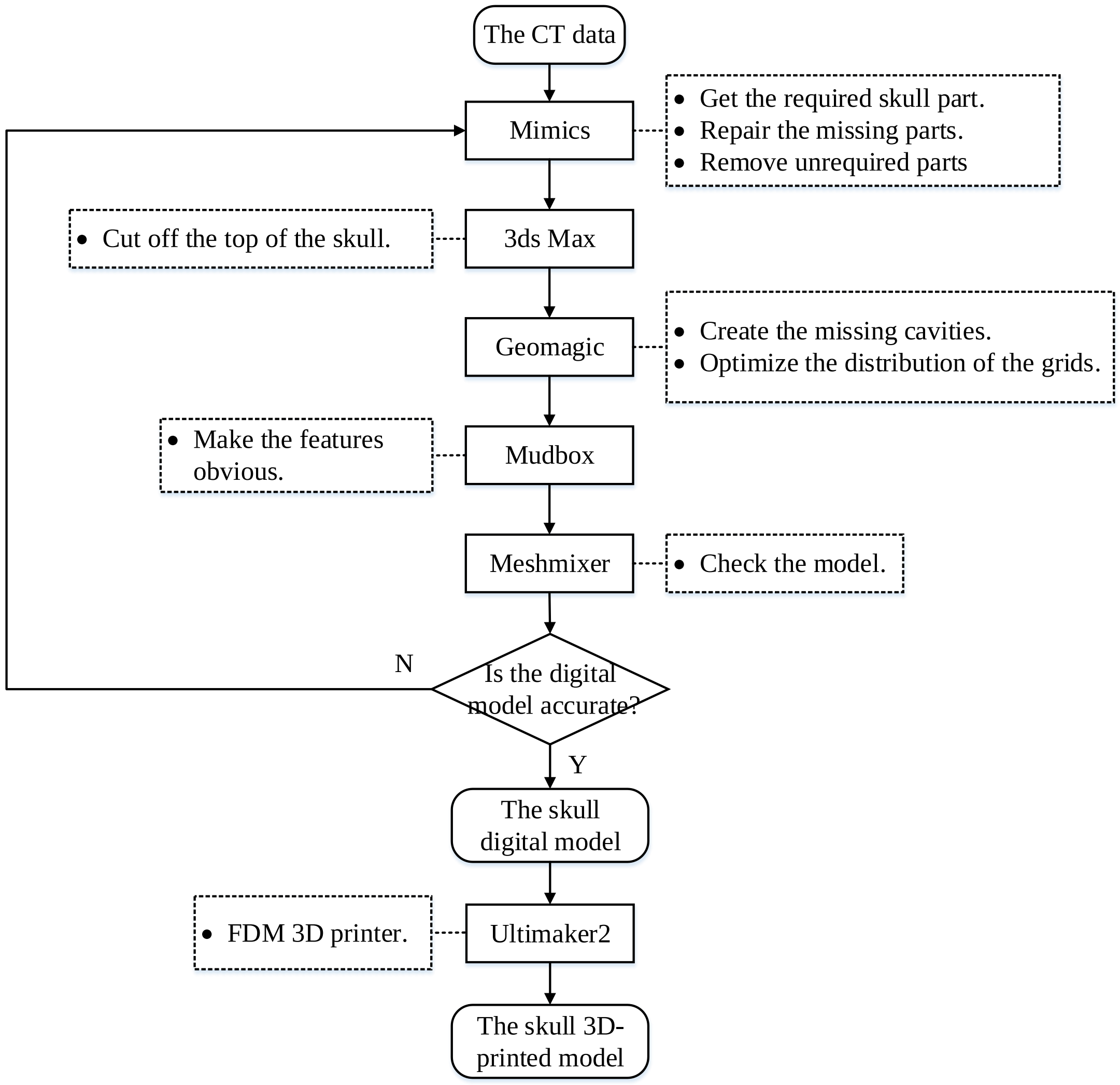

Fig.14 Summary of workflow. We use five 3D manipulation tools to complete this procedure. We use Mimics to read the data and repair the model by adding the missing parts, use 3ds Max to cut off the top part of the skull use Geomagic to fix holes in the surface and automatically repair the imperfections in a polygon mesh, use Mudbox to make the features more obvious, and use Meshmixer to check the model. Then the model is fixed and ready for printing.

Firstly, Mimics can read the DICOM files directly and output the STL file and meanwhile help edit the 3D digital data. Many tools in Mimics, such as setting thresholding value and editing mask in 3D, are useful in repairing the structure as we have shown. Secondly, the 3ds Max is a professional 3D data processing tool and the “slice” tool is useful in removing the skull cap and we can quickly select some holes to delete or repair. Thirdly, Geomagic Studio is used to repair the surface holes in the study. We can repair the data depending on the distribution of the meshes and also improve the distribution to make the surface of the model better. Then, Mudbox is a software used in the digital carving. We use it to make some features more obvious, such as shaping the plasticine. Before printing, Meshmixer can help us check the digital model. Since the anatomy structure is complicated, we have to use different tools to check and repair the digital model to make sure the model can be printed well.

The printed model has already been used in a randomized controlled trial (RCT), comparing the learning efficiency of 3D printed skulls with that of cadaveric skulls and atlas. The result shows that the 3D printed skull model has advantages in assisting anatomy study, especially in structure recognition, compared with traditional education tools [8].

The procedures shown in the paper can be applied for other samples. Besides, according to the doctors' or lecturers' requirements, the model can be correspondingly designed or deformed during the procedures. Since the 3D printing technology is becoming more and more developed, and the building volume of the printer is various and the procedure is efficient, not only the skull model but also other parts or organ models of the body can be 3D-printed and applied in the medical education. However, the procedure presented in this study requires the combination of medical knowledge and computer skills. It is not easy for a surgeon or an engineer to complete the entire project independently. In this case, they have to keep in touch during the whole procedure, which to some extent would slow the progress.

Apart from the medical education, the application of the 3D printing technology and computer-assisted medical image processing can also be used in the clinical medicine. The surgeries, especially endoscopic ones, require the surgeons to have excellent surgery skills and adequate knowledge about the body structures. However, the endoscopic checking is not enough for the needed information. In this case, we can scan the patient's sick part and 3D-print it. This is helpful in the research and analysis of the disease, and the surgery simulation.

## Conclusion

This kind of skull models can be produced and designed repeatedly with a low cost and high efficiency. The procedures can be applied in the manufacturing of other medical models. The application of the 3D printing technology is worth developing in the field of medical education.

## Compliance with ethical standards

**Conflict of interest** The authors declare that they have no conflict of interest.

**Ethical approval** The local ethics committee considered that this study had been carried out in accordance with the Declaration of Helsinki.

**Informed consent**

Informed consent was obtained from all individual participants included in the study.

## Reference

1. Schmitt B, Wacker C, Ikemoto L, Meyers FJ, Pomeroy C (2014) A transparent oversight policy for human anatomical specimen management: the University of California, Davis experience. Academic Medicine Journal of the Association of American Medical Colleges 89(3): 410-4. doi: 10.1097/ACM.0000000000000135.
2. Gunderman RB (2010) Giving ourselves: the ethics of anatomical donation. Anatomical Sciences Education 1(5):217-219. doi: 10.1002/ase.39.
3. Tabinda H (2011) Is dissection humane? J Med Ethics Hist Med 4: 4.

4. Waran V, Menon R, Pancharatnam D, Rathinam AK, Balakrishnan YK, Tung TS, et al. (2012) The creation and verification of cranial models using three-dimensional rapid prototyping technology in field of transnasal sphenoid endoscopy. American Journal of Rhinology & Allergy 26(5):132-136. doi: 10.1055/s-0031-1300964.
5. Lee SJ, Lee HP, Tse KM, Cheong EC, Lim SP (2012) Computer-aided design and rapid prototyping-assisted contouring of costal cartilage graft for facial reconstructive surgery. Craniomaxillofacial Trauma & Reconstruction 5(2): 75-82. doi: 10.1055/s-0031-1300964.
6. Matthew DT, Stephen D, Duncan B (2012) 3-D printout of a DICOM file to aid surgical planning in a 6 year old patient with a large scapular osteochondroma complicating congenital diaphyseal aclasia. J Radiol Case Rep 6(1): 31-37. doi: 10.3941/jrcr.v6i1.889.
7. Debarre E, Hivart P, Baranski D, Déprez P (2012) Speedy skeletal prototype production to help diagnosis in orthopaedic and trauma surgery. Methodology and examples of clinical applications. Orthopaedics & Traumatology Surgery & Research Otsr 98(5): 597. doi: 10.1016/j.otsr.2012.03.016.
8. Chen S, Pan Z, Wu Y, Gu Z, Li M, Liang Z, Zhu H, Yao Y, Shui W, Shen Z, Zhao J. (2017) The role of three-dimensional printed models of skull in anatomy education: a randomized controlled trail. Scientific Reports 7(1): 575. doi:10.1038/s41598-017-00647-1.
9. Shui W, Zhou M, Chen S, Pan Z, Deng Q, Yao Y, Pan H, He T, Wang X (2017) The production of digital and printed resources from multiple modalities using visualization and three-dimensional printing techniques. International journal of computer assisted radiology and surgery, 12(1): 13-23.
10. Li Z, Li Z, Xu R, Li M, Li J, Liu Y, Sui D, Zhang W, Chen Z (2015) Three-dimensional printing models improve understanding of spinal fracture–a randomized controlled study in China. Sci Rep 5:11570. doi:10.1038/srep11570.
11. Wagner JD, Baack B, Brown GA, Kelly J (2004) Rapid 3-dimensional prototyping for surgical repair of maxillofacial fractures: a technical note. J Oral Maxillofac Surg 62(7):898–901. doi:10.1016/j.joms.2003.10.011
12. Kumar S et al. Reinforcement of stereolithographic resins for rapid prototyping with cellulose nanocrystals. ACS Appl Mater Interfaces. 2012;4(10):5399–407. doi:10.1021/am301321v.
13. Friedman T, Michalski M, Goodman TR, Brown JE. 3D printing from diagnostic images: a radiologist's primer with an emphasis on musculoskeletal imaging—putting the 3D printing of pathology into the hands of every physician. Skeletal radiology. 2016 Mar 1;45(3):307-21.
14. Sectra Inc. 2017, Imaging IT solutions that lead the way in customer satisfaction. website: https://www.sectra.com/medical/orthopaedics/solutions/planning_tools_3d/

**Acknowledgement**

We would like to acknowledge support in part from the National Natural Science Foundation of China under Grants 61773382, 61773381, 61533019, Chinese Guangdong's S&T project (2015B010103001, 2016B090910001), Dongguan's Innovation Talents Project (Gang Xiong), Capital Characteristic Clinic Project (No. Z141107002514051), Young Teacher Training Program (No, X103852), Education Reform Program (No. X103851).

## Author Contributions

Z.S. and Y.X. wrote the paper. Y.Y., C.G., G.X. and X.D. conducted the experiment and analyzed the results. X.S. and Y.L. constructed the model. Z.P. and S.C. collected data. H.P. reviewed the paper. All authors reviewed the manuscript.